\documentclass[showpacs,amsmath,nofootinbib,showpacs]{revtex4} 
\usepackage{graphicx}
\usepackage{epsfig}

\usepackage{graphicx}
\usepackage{dcolumn}
\usepackage{bm}

\begin{document}

\title{Contributions of dark matter annihilation within ultracompact minihalos to the 21cm background signal}

\author{Yupeng Yang$^{1,2}$} \email{aynuyyp@163.com}

\affiliation{$^1$Collage of Physics and Electrical Engineering, Anyang Normal University, Anyang, 455000, China\\
$^2$Joint Center for Particle, Nuclear Physics and Cosmology, Nanjing, 210093, China}

\begin{abstract}
In the dark age of the universe, any exotic sources, e.g. the dark matter annihilation, 
which inject the energy into the intergalactic medium (IGM) 
will left some imprint on the 21cm background signal. Recently, one new kind of 
dark matter structure named ultracompact dark matter minihalo (UCMHs) was proposed. 
Near the inner part UCMHs, the distribution of dark matter particles are steeper than that of the general 
dark matter halos, $\rho_{\rm UCMHs}(r) \sim r^{-2.25}$, and the formation time of UCMHs is earlier, $z_c \sim 1000$. 
Therefore, it is excepted that the dark matter annihilation within UCMHs 
can effect the 21cm background signal. In this paper, we investigated the contributions of 
the dark matter annihilation within UCMHs to the 21cm background signal. 
\end{abstract}


\maketitle
\section {Introduction}
In the earlier epoch, the Universe was in the fully ionized phase. 
With the expansion of the Universe, the temperature of IGM decreased and protons combined 
with electrons to form hydrogens. This process is named recombination and 
occurs at the redshift $z\sim1100$. After the recombination, the Universe 
went into the epoch named "dark age". The most 
exciting way of detecting the "dark age" is to observe the 21cm background signal. The 21cm signal is 
caused by the transition of hyperfine split ($n=1$) of the hydrogen. It is the 
result of the competition among the temperature of baryonic gas ($T_{k}$), radiation ($T_{\gamma}$) 
and spin ($T_s$). Therefore, the energy injection during 
the dark age will affect $T_k$ and $T_s$, and left imprint on the 21cm background signal~\cite{21cm_review_1,21cm_review_2}. 

As a kind of extra source, dark matter annihilation can also affect the 21cm signal.  
The basic idea is that the dark matter annihilation productions, 
e.g. photons ($\gamma$), electrons ($e^-$) and 
positrons ($e^+$), have interactions with the particles which are present 
in the Universe causing the heating, ionization and excitation of IGM~\cite{xlchen,lezhang}. Therefore, the temperature 
$T_k$ and $T_s$ will be affected and changed. 
Due to the factor that dark matter annihilation rate is proportional to the 
number density square of particles, the influence of dark matter annihilation on IGM can be studied 
using the 21cm background signal~\cite{valdes,deltaTb,Natarajan}. 
Recently, a new kind of dark matter 
structure named ultracompact dark matter minihalos (UCMHs) was proposed~\cite{ucmhs_1}. They can 
be formed in the early Universe via the collapse of large density 
perturbations, $0.001 \lesssim \delta \rho/\rho \lesssim 0.3$. 
Compared with the general dark matter halos, 
the density profile of UCMHs is steeper, $\rho(r) \sim r^{-2.25}$, 
and the formation time is earlier, $z\sim1000$. Therefore, it is excepted that 
the dark matter annihilation rate is large within UCMHs. Several works 
have discussed the gamma-ray flux and neutrino flux from UCMHs due to 
the dark matter annihilation or decay~\cite{scott_prl,prd_neutrino,jcap,epl,ijmpa,raa}. 
Due to the large annihilation rate of dark matter within UCMHs, 
the gamma-ray or neutrino flux from UCMHs can exceed the threshold of some detectors or 
the atmosphere neutrino background. In Refs.~\cite{prd_1,epjplus,dongzhang}, the authors investigated the 
effects of UCMHs on the anisotropy of cosmic microwave background and 
the structure formation due to the dark matter annihilation. 
In this paper, we will focus on the 
contributions of UCMHs to the 21cm background signal. 

Dark matter as an essential component of the Universe has been confirmed 
while its nature is still unknown. At present, the mostly researched model is 
the weakly interactive massive particles (WIMPs). 
The typical mass of WIMPs is $m_{\chi} \sim 100$ GeV $-10$ TeV, and the thermally 
averaged cross section is $\left<\sigma v\right> \sim 3 \times 10^{-26} \mathrm{cm^{3} s^{-1}}$~\cite{dm_1,dm_2}. 
However, in order to explain 
the gamma-ray excess of the Milky Way center the dark matter particle mass would be $m_{\chi} \sim 10$ GeV, e.g. Ref.~\cite{Hooper}. 
The observations on the positrons and electrons of the cosmic ray 
imply that the dark matter particle mass is $m_{\chi} \sim 1$ TeV  
and the thermally averaged cross section is $\left<\sigma v\right> \sim 10^{-23} \mathrm{cm^{3} s^{-1}}$, e.g. see the review~\cite{xjb}. 

In Ref.~\cite{PhysRevLett.115.231301}, after the analysis of $\gamma$-ray 
data of the Milky Way 
dSphs from the Fermi-LAT the authors excluded the thermally averaged cross section 
$\left<\sigma v\right> \sim 2.2 \times 10^{-26} \mathrm{cm^{3} s^{-1}}$ 
for the dark matter particle mass $m_{\chi} < 100~\rm GeV$. 
The observational results of H.E.S.S towards the Galactic center 
for the very high energy $\gamma$-ray flux 
show that there is no a residual $\gamma$-ray flux in the energy range between 
$300~\rm GeV$ and $30~\rm TeV$~\cite{PhysRevLett.106.161301}, and the 
thermally averaged cross section which is larger than 
$3\times 10^{-23} \mathrm{cm^{3} s^{-1}}$ is excluded 
for the Einasto density profile.

In this paper, we will focus on the general case, 
$m_{\chi} \sim 100$ GeV with $\left<\sigma v\right> 
\sim 3 \times 10^{-26} \rm cm^{3}s^{-1}$, and our results can be applied 
for the other cases. 
For the cosmological parameters, we used the results given by the Planck collaboration~\cite{planck}.  

This paper is organized as follows: In Sec. II, we briefly 
show the general picture of the 21cm background and UCMHs. 
In Sec. III, we calculate the impact of dark matter annihilation within UCMHs on the 21cm background. 
The main conclusions and discussions are given in Sec. IV.

\section {The general picture of 21cm background and UCMHs} 
\subsection{The 21cm background from the dark age}

The ground state of hydrogen ($n=1$) can split into triplet and singlet states, 
which is named hyperfine structure. The energy change of these two levels 
is $E = 5.9 \times 10^{-6}$ eV which corresponds to the wavelength of photon 
$\lambda = 21$ cm. The transition between two states is usually expressed in 
the form of spin temperature, $T_s$, which is defined as

\begin{equation}
\frac{n_1}{n_0} = 3\mathrm{exp}\left(-\frac{T_\star}{T_s}\right),
\end{equation}
where $n_1$ and $n_0$ are the number density of hydrogens in triplet and singlet 
states, $T_{\star} = 0.068$ K is the equivalent temperature corresponding to the transition energy. During the 
evolution of the universe, there are mainly three processes which have influence on the spin 
temperature (i) the background photons, e.g. comic microwave background (CMB), 
which can be absorbed by hydrogen atoms; (ii) the collisions of the hydrogen atoms 
with other particles, such as hydrogen atoms, electrons and protons; 
(iii) The resonant scattering of Ly$\rm \alpha$ photons which is 
named the Wouthuysen-Field effect. Including these effects, the spin temperature 
can be written as

\begin{equation}
T_s = \frac{T_{\rm CMB}+(y_\alpha + y_c)T_k}{1+y_\alpha+y_c},
\label{eq:main}
\end{equation}
where $y_{\alpha}$ corresponds to the 
Wouthuysen-Field effect, and in this work we use the form as~\cite{binyue}

\begin{equation}
y_\mathrm{\alpha} = \frac{P_{10}T_{\star}}{A_{10}T_\mathrm{k}}e^{-0.3\times(1+z)^{0.5}
T_\mathrm{k}^{-2/3}\left(1+\frac{0.4}{T_\mathrm{k}}\right)^{-1}},
\end{equation}
where $A_{10} = 2.85 \times 10^{-15} s^{-1}$ 
is the Einstein coefficient of the hyperfine spontaneous transition. 
$P_{10}$ is the de-excitation rate of 
the hyperfine triplet state due to Ly$\rm \alpha$ scattering, here we use the 
form $P_{10} = 1.3 \times 10^{9} J_{\alpha}$, $J_{\alpha}$ is the intensity of 
Ly$\rm \alpha$ radiation~\cite{j_alpha}

\begin{equation}
J_\alpha (z) = \frac{n_{H}^2hc}{4\pi H(z)}\left[x_{e}x_{p}\alpha_{2^2P}^\mathrm{eff}+x_{e}x_\mathrm{HI}\gamma_\mathrm{eH}
		+\frac{\chi_{\alpha}E_\mathrm{UCMHs}}{n_{H}h\nu_{\alpha}}\right],
\label{J_alpha}
\end{equation}
where $x_e$ is the fraction of the free electrons defined as $x_{e} = \frac{n_{e}}{n_{e}+n_{H}}$, 
$x_p = x_e$ and $x_\mathrm{HI} = 1 - x_e$ are the fraction of protons and hydrogens respectively. 
$\alpha_{2^2P}^\mathrm{eff}$ is the effective recombination coefficient to the $2^2P$ level and 
we adopt the form as~\cite{Ripamonti} 

\begin{equation}
\alpha_{2^2P}^\mathrm{eff} = 1.67 \times 10^{-13}\left(\frac{T}{10^4K}\right)^{-0.91-\frac{2}{75}\mathrm{log_{2}}\frac{T}{10^4K}} 
\ [\mathrm{cm^{3} s^{-1}}].
\end{equation}
$\gamma_\mathrm{eH}$ is the excitation rate of hydrogen due to the collisions of electrons~\cite{Ripamonti}, 

\begin{equation}
\gamma_\mathrm{eH} = 2.2 \times 10^{-8}e^{-118400K/T_{k}} \ [\mathrm{cm^{3}s^{-1}}].
\end{equation}

The third term of Eq.(\ref{J_alpha}) is the effect of dark matter annihilation 
within UCMHs. $\chi_\alpha = (1-x_e)/6$ and $E_\mathrm{UCMHs}$ is the energy injection 
rate of dark matter annihilation, and it will be given in the next section.  

In the Eq.(\ref{eq:main}), $y_c$ corresponds to the collision effect between 
hydrogen, electrons and protons, it can be written as (see e.g. the Ref.~\cite{Kuhlen} )

\begin{equation}
y_c = \frac{(C_{\rm HH}+C_{\rm eH}+C_{\rm pH})T_\star}{A_{10}T_K}, 
\end{equation}
where $C_{\rm HH,eH,pH}$ are the de-excitation rate, and we use the form as~\cite{Kuhlen,Liszt,deltaTb} 

\begin{equation}
C_\mathrm{HH}  = 3.1 \times \rm 10^{-11}T_{k}^{0.357}e^{-32K/T_{K}} \times n_\mathrm{HI} \ [\mathrm{cm^{3}s^{-1}}]
\end{equation}

\begin{equation}
C_\mathrm{eH}  = 10^{-9.607+0.5\rm log(T_{k})e^{-(log(T_{k}))^{4.5}/1800}} \times n_\mathrm{e} \ [\mathrm{cm^{3}s^{-1}}] 
\end{equation}

Following the previous discussions (see e.g. the Ref.~\cite{deltaTb}), the term of $C_\mathrm{pH}$ can be neglected safely 
due to the slight effect. 

For the observations of 21cm background, the mostly used 
quantity is the brightness temperature, $\delta T_b$, which is 
defined as the differences between the spin temperature and the CMB temperature~\cite{deltaTb,Ciardi}, 

\begin{equation}
\delta T_b \simeq 26 \times (1-x_e)\left(\frac{\Omega_{b}h}{0.02}\right)\left[\frac{1+z}{10}\times\frac{0.3}{\Omega_m}\right]^{\frac{1}{2}}
\left(1-\frac{T_{\rm CMB}}{T_s}\right){\rm mK}.
\label{delta_tb}
\end{equation}

In the section, we reviewed the basic quantity of the 21cm background simply, 
and for more detail discussions one can see e.g. the review~\cite{review_1,review_2}.

\subsection{The basic character of UCMHs}

It is well known that in the early Universe the 
density perturbations with the amplitude $\delta \rho /\rho \sim 10^{-5}$ 
are the seeds of the present cosmological structures. 
If the density perturbations are larger than $\sim 0.3$ the primordial black holes 
can be formed. Recently, the authors of~\cite{ucmhs_1} proposed that a new 
kind of dark matter structures named ultracompact minihalos can be formed if 
the amplitude of density perturbations lie between the above mentioned values. The density profile 
of UCMHs can be obtained from via simulation, 

\begin{equation}
\rho_{\rm UCMHs}(r,z) = \frac{3f_{\chi}M_{\rm UCMHs}(z)}{16\pi 
R_{\rm UCMHs}(z)^{0.75}r^{2.25}}, 
\label{rho_ucmhs}
\end{equation}
where $f_{\chi} = \Omega_{\rm DM}/(\Omega_b + \Omega_{\rm DM}) = 0.83$~\cite{scott_prl}. 
$R_{\rm UCMHs}(z) = 0.019\left(\frac{1000}{1+z}
\right)\left(\frac{M_{\rm UCMHs(z)}}{M_{\odot}}\right)^{1/3} \mathrm{pc}$ is the radius 
of UCMHs at the redshift $z$. $M_{\rm UCMHs}(z)$ is the mass of UCMHs at the 
redshift $z$, and it is related with the initial mass $M_i$ contained within 
the perturbation scale as they entering the horizon~\cite{scott_prl}, 

\begin{equation}
M_{\rm UCMHs}(z) = M_i \left(\frac{1+z_{\rm eq}}{1+z}\right),
\end{equation}  
where $z_{\rm eq} \simeq 3160$ is the redshift at which the energy density of radiation and 
matter are equal.

From Eq.~(\ref{rho_ucmhs}) it can be seen that for $r \to 0$ the density 
$\rho \to \infty$. In order to avoid this divergence we truncate the density 
profile at the radius $r_{\rm min}$ and it satisfies the condition as~\cite{rcut,scott_prl} 

\begin{equation}
\rho_{\rm UCMHs}(r_{\rm min}) = \frac{m_{\chi}}{\langle \sigma v \rangle 
(t-t_i)},
\end{equation}
where $t_i$ is the formation time of UCMHs. This relation is the result of considering 
the dark matter annihilation with the halos. For smaller radius $r < r_{\rm min}$, 
we set the density is constant, $\rho_{\rm UCMHs}(r<r_{\rm min}) = 
\rho_{\rm UCMHs}(r_{\rm min})$.

\section{The imprint of dark matter annihilation within UCMHs on the 21cm background}

The influences of dark matter annihilation on the evolution of the Universe have been researched 
by many works. The main effects are on the evolution of the ionization degree and the 
temperature of IGM. Including the dark matter annihilation, the change of 
the ionization degree ($x_e$) and the temperature of IGM ($T_b$) with the time can be written as~\cite{lezhang,xlchen} 
 

\begin{equation}
(1+z)\frac{dx_e}{dz}=\frac{1}{H(z)}[R_{s}(z)-I_{s}(z)-I_{\chi}(z)],
\end{equation}

\begin{eqnarray}
(1+z)\frac{dT_b}{dz} =&&\frac{8\sigma_{T}a_{R}T_{\rm cmb}^4}{3m_{e}cH(z)}
\frac{x_e}{1+f_{\rm He}+x_e}(T_{b}-T_{\rm cmb}) \nonumber \\
&&-\frac{2}{3k_{B}H(z)}\frac{K_{\chi}}{1+f_{\rm He}+x_e}+T_b,
\end{eqnarray}
where $R_{s}(z)$ and $I_{s}(z)$ are the standard recombination rate and 
ionization rate caused by the standard sources, respectively. $I_{\chi}$ and $K_{\chi}$ 
are the ionization 
rate and heating rate associated with the dark matter annihilation,


\begin{equation}
I_{\chi} = \chi_{i}f\frac{2m_{\chi}c^2}{n_{b}E_{b}}\Gamma_{\rm DM}
\label{ion}
\end{equation}
\begin{equation}
K_{\chi} = \chi_{h}f\frac{2m_{\chi}c^2}{n_{b}E_{b}}\Gamma_{\rm DM}
\label{heat}
\end{equation}
where $f$ is the energy fraction which is injected into the IGM from dark matter annihilation. It depends 
on the redshift and the annihilation channel. In fact, for the dark matter annihilation, 
$f$ depends on the redshift slightly~\cite{j_alpha}. 
In this paper, we adopted $f = 1$ for our calculations and very detailed discussions 
will be presented in the future work. 
$\chi_i$ and $\chi_h$ are the ionizing and heating fractions 
of energy which deposit into the IGM. 
There are several parameterizations of this form. In this paper, 
we adopted the widely used forms proposed by the authors of~\cite{xlchen}, $\chi_i = (1-x_e)/3$ and $\chi_h = (1+2x_e)/3$. 
In this work, we considered the dark matter annihilation within UCMHs, 
so the annihilation rate $\Gamma_{\rm DM}$ in Eqs. (\ref{ion}) and (\ref{heat}) can be written as~\cite{prd_1}, 

\begin{eqnarray}
\Gamma_{\rm UCMHs} =&&f_{\rm UCMHs} \frac{\rho_{\rm 0,c}}{M_{\rm 0,UCMHs}}(1+z)^3 \nonumber \\
&&\times \frac{\langle\sigma v\rangle}{m_{\chi}^2} \int 4\pi r^2 \rho_{\rm UCMHs}(r,z)^2 dr 
\end{eqnarray}
where $f_{\rm UCMHs} = \rho_{\rm UCMHs}/\rho_{0,c}$, $\rho_{\rm 0,c}$ is the current 
critical density of the universe. Be silimar to the PBHs case, here, we have assumed that the mass of UCMHs is the same 
when they are formed, and this assumption means 
that the mass function of UCMHs is in the delta form~\cite{prd_1,scott_prl},  $dn/dM \sim \delta (M-M_{\rm UCMHs})$. 
We also assume that there are no mergers between of them. Now, we can write the energy injection rate 
in Eq.(\ref{J_alpha}) as 

\begin{equation}
E_{\rm UCMHs} = 2m_{\chi}\frac{\Gamma_{\rm UCMHs}}{n_{b,0}}
\end{equation}
where $n_{b,0}$ is current number density of baryon. 
We modified the public code {\small\bf RECFAST} in the {\small \bf CAMB} \footnote{http://camb.info/} to include the effect of dark matter annihilation within UCMHs, 
and the evolution of $x_e$ and $T_{\rm IGM}$ can be obtained using the modified code. 
In this section, we will not consider the contributions of general dark matter halos, however, the relative discussions 
will be given in the next section.

\begin{figure}
\epsfig{file=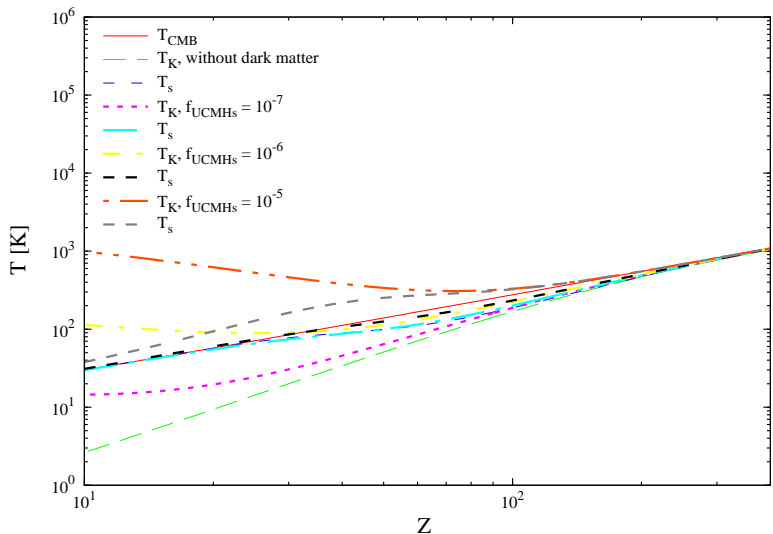,width=0.6\textwidth}
\caption{The evolution of temperature, $T_{\rm IGM}$ and $T_s$, with and without UCMHs. Here we set the parameters of 
dark matter as: $m_{\chi} = 10 \rm GeV$, $\langle\sigma v\rangle = 3 \times 10^{-26} \rm cm^{3}s^{-1}$. 
The fraction of UCMHs are $f_{\rm UCMHs} = 10^{-7}, 10^{-6}$ and 
$10^{-5}$. The evolution of CMB temperature $T_{\rm CMB}$ 
is also plotted (thin solid red curve).}
\label{fig_tem}
\end{figure}

\begin{figure}
\epsfig{file=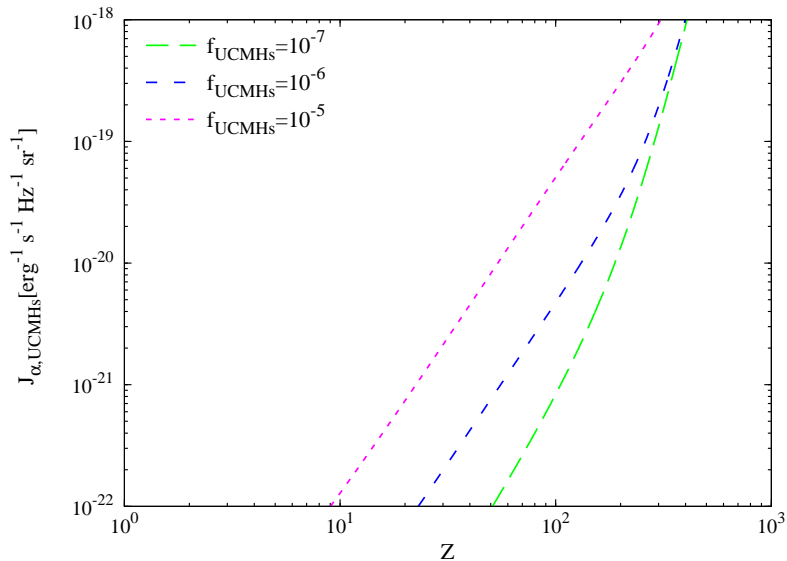,width=0.6\textwidth}
\caption{The Ly$\alpha$ intensity from UCMHs due to the dark matter annihilation. 
The solid red line and the dashed green line correspond to the 
$f_{\rm UCMHs}= 10^{-7}, 10^{-6}$ and  $10^{-5}$, respectively. The other parameters are 
the same as in Fig.~\ref{fig_tem}.}
\label{lyalpha}
\end{figure}

In Fig.~\ref{fig_tem}, the changes of $T_{k}$ and $T_{s}$ with redshift 
for different cases are shown. For comparison, we also plotted the case without 
dark matter. The thin red solid line represents 
the temperature of CMB, $T_{\rm CMB}$. The parameters of dark matter particle are $m_{\chi} = 100$GeV, 
$\langle \sigma v \rangle = 3 \times 10^{-26} \rm cm^{3}s^{-1}$. From the figure, 
it can be seen that the temperature of IGM becomes notably different compared with the no dark matter case. 
The $T_\mathrm{IGM}$ is higher for the 
larger fraction of UCMHs. The spin temperature $T_{s}$ also show the similar 
variation trend. We also show the evolution of intensity of the Ly$\alpha$ due to dark matter annihilation within UCMHs 
in Fig.~\ref{lyalpha}. In this figure, one can see that the UCMHs provide 
extra sources of Ly$\alpha$ during the early time. At the redshift $z \sim 300$, 
for $f_{\rm UCMHs} = 10^{-5}$, the intensity of Ly$\alpha$ is $J_{\alpha} \sim 10^{18} 
\rm erg \ s^{-1} Hz^{-1} sr^{-1}$. In Ref.~\cite{binyue}, the authors 
showed the contributions from the first stars to the Ly$\alpha$ background (Fig.5 of that Ref.). These 
contributions increase dramatically only after $z \sim 30$, and the intensity of Ly$\alpha$ reach the peak value $J_{\alpha} 
\sim 10^{-21} \rm erg \ s^{-1} Hz^{-1} sr^{-1}$ at redshift $z \sim 10$.

\begin{figure}
\epsfig{file=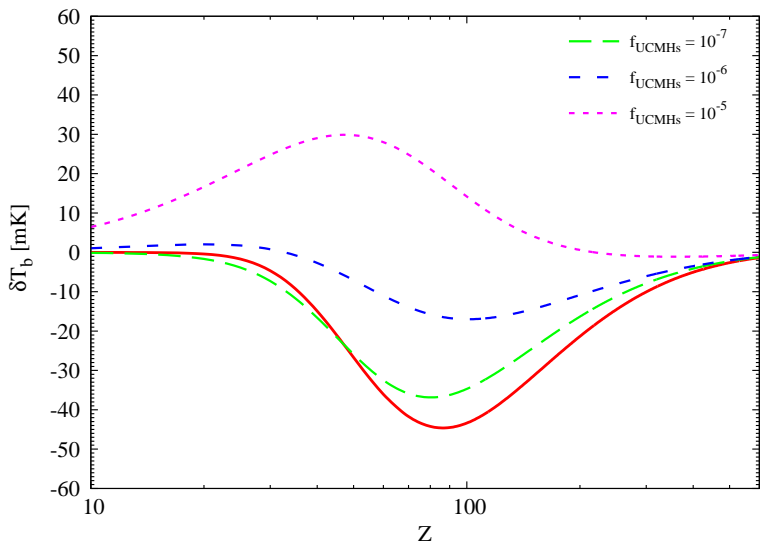,width=0.6\textwidth}
\caption{The evolution of the differential brightness 
temperature $\delta T_{b}$. The solid red line is the case without UCMHs. 
The long green dashed and the short blue dashed lines correspond 
to the $f_{\rm UCMHs} = 10^{-7}, 10^{-6}$ and $10^{-5}$ from bottom to up, 
respectively.}
\label{fig:delta_tb}
\end{figure}

\begin{figure}
\epsfig{file=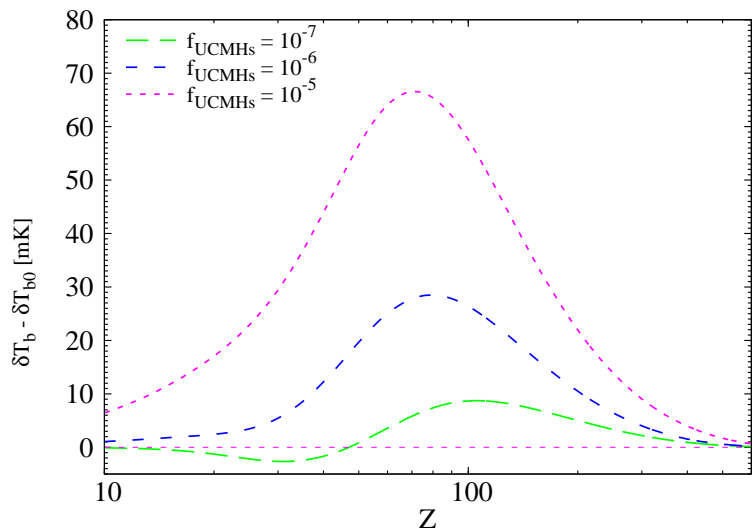,width=0.6\textwidth}
\caption{The differences of the differential brightness temperature between 
the cases of with UCMHs and without UCMHs. The line style and parameters are 
the same as Fig.~\ref{fig:delta_tb}.}
\label{fig:com}
\end{figure}

Having obtained the evolution of spin temperature $T_s$, we can now calculate the differential brightness 
temperature $\delta T_b$ using Eq.~\ref{delta_tb}. 
The results are shown in Fig.~\ref{fig:delta_tb}. In this plot, we also 
show the case of without UCMHs (red solid line) for comparison. 
For the smaller fraction of UCMHs, $f_{\rm UCMHs} \lesssim 10^{-6}$, there are absorption features in $\delta T_b$. 
If the fraction of UCMHs is larger than $10^{-6}$, the emission feature appear. 
We also plotted the 
differences between the cases of with and without UCMHs, $\Delta \delta T_{b} \equiv \delta T_{b} - \delta T_{b,0}$, 
in Fig.~\ref{fig:com}. As shown in this figure, the largest differences are $\Delta \delta T_{b} \sim  67 {\rm mk}$ 
at redshift $z \sim 67$ for $f_{\rm UCMHs} = 10^{-5}$, 
$\Delta \delta T_{b} \sim  28 {\rm mk}$ at redshift $z \sim 80$ for $f_{\rm UCMHs} = 10^{-6}$ and 
$\Delta \delta T_{b} \sim  8.7 {\rm mk}$ at redshift $z \sim 100$ for $f_{\rm UCMHs} = 10^{-7}$, respectively.

\begin{figure}
\epsfig{file=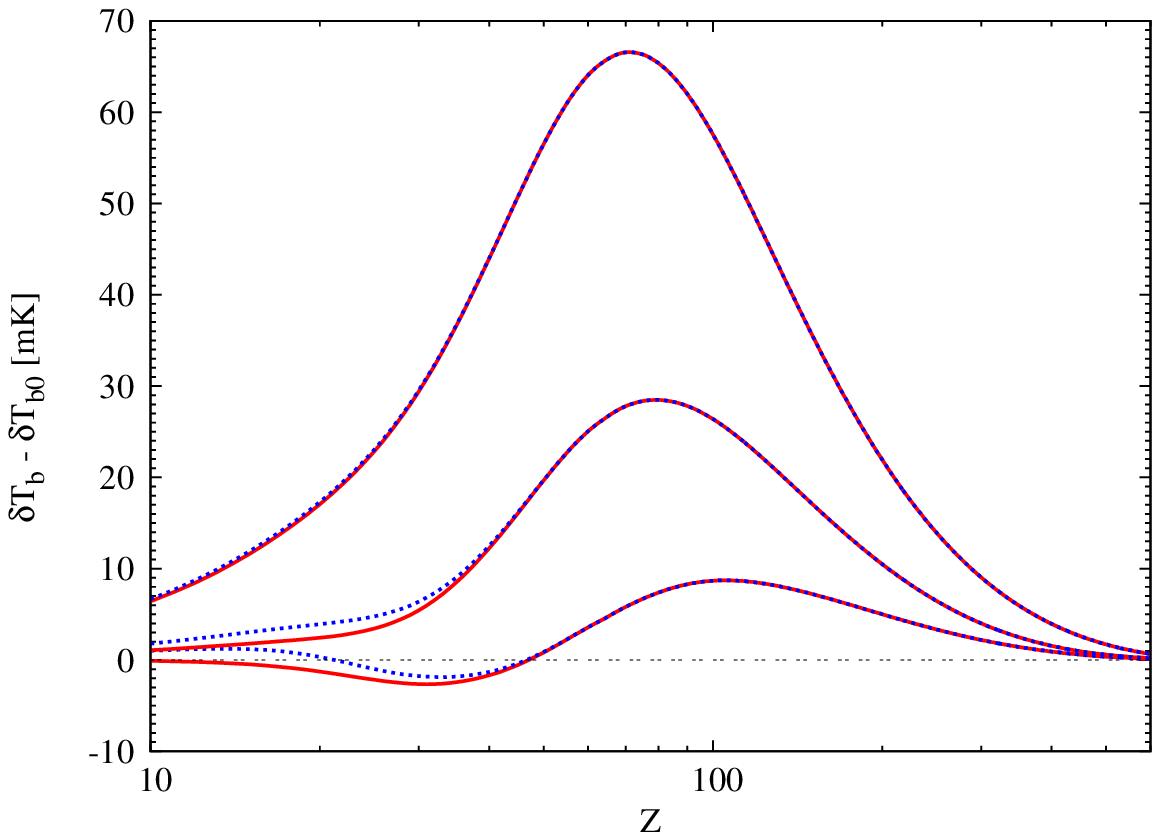,width=0.6\textwidth}
\caption{The evolution of $\Delta \delta T_{b} \equiv \delta T_{b} - \delta T_{b,0}$ 
for different fraction of UCMHs 
with (dotted blue lines) and without (solid red line) general dark matter halos. 
From the bottom to up, the fraction of UCMHs are $f_{\rm UCMHs} = 10^{-7}, 
10^{-6}$ and $10^{-5}$. The parameters of dark matter are the same as the previous 
figures.}
\label{fig:com_stru_ucmhs}
\end{figure}


\section{discussions}

We have investigated the effects of dark matter annihilation within 
UCMHs on the 21cm background signal. The formation time of UCMHs is early ($z \sim 1000$) 
and the density profile is steep ($\rho(r) \sim r^{-2.25}$). So the dark matter annihilation 
rate within UCMHs is higher than that of general dark matter halos, e.g. NFW models. 
Due to the extra energy injection from UCMHs, the evolution of the temperature of IGM and 
spin with the redshift will be changed. $T_{\rm K,IGM}$ can be up to $1000$K at the redshift $z \sim 10$ for 
$f_{\rm UCMHs} = 10^{-5}$. The spin temperature 
is also changed obviously at the redshift $z \sim 200$ especially for large fraction of UCMHs. 
Moreover, UCMHs can also provide extra Ly$\alpha$ background during the early epoch. 
For the observation of the 21cm background signal the mostly used quantity 
is the differential brightness temperature $\delta T_b$. We found that for the 
small fraction of UCMHs, $f_{\rm UCMHs} \lesssim 10^{-6}$, there is an fabsorption feature, and 
the emission feature appears for large fraction of UCMHs. We also investigated the quantity 
$\Delta \delta T_{b}$ which is defined as 
$\Delta \delta T_{b} \equiv \delta T_{b} - \delta T_{b0}$. 
We found that the values of $\Delta \delta T_{b}$ are changed for different 
fraction of UCMHs. For the fraction of UCMHs $f_{\rm UCMHs} = 10^{-5}$, 
$\Delta \delta T_{b}$ can be up to $\sim 67\rm mk$ at the redshift $z \sim 67$, 
and it becomes smaller with the decreasing of the fraction of UCMHs.  

Except for UCMHs the general dark matter halos have also contributions to the 
21cm background signal. The effects of general dark matter halos can be 
treated as a 'clumping factor' $C(z)$ relative to the smooth case~\cite{deltaTb}. 
According to the simulation, there are many subhaols and sub-sub halos.  
In this paper, we considered the subhalos while neglecting the sub-subhalos. We adopted 
the smallest mass of halos as $\sim 10^{-6} \rm M_{\odot}$~\cite{dmhalos}. The halos mass 
within subhalos is about $\sim 10\%$, and we used the power law form of mass 
function $\sim M^{-\beta}$ with $\beta = 1.95$~\cite{subhalos}. We used the NFW dark matter 
halo model for our calculations. For different fraction of UCMHs, $f_{\rm UCMHs} = 10^{-7}, 10^{-6}$ and $10^{-5}$, 
$\Delta \delta T_{b}$ are shown in Fig.~\ref{fig:com_stru_ucmhs}. 
The obvious differences between the contributions of UCMHs and general dark matter halos 
to the 21cm background signal appear at the redshift $z \sim 40$ for the small fraction of UCMHs, $f_{\rm UCMHs} < 10^{-5}$. 
For large fraction of UCMHs, $f_{\rm UCMHs} \gtrsim 10^{-5}$, the main contributions are from UCMHs. 
For the observations of 21cm background signals, as shown in Fig.~\ref{fig:com}, the values of $\Delta \delta T_{b}$ can reach 
$\sim 27 \rm mk$ at redshift $z \sim 30$. Therefore, in order to find the 
impacts of UCMHs on IGM due to the dark matter annihilation 
the systematics of experiments e.g. EDGES(Experiment for 
Detecting
the Global EOR Signature)~\cite{0004-637X-676-1-1} should be below $27\rm mk$ at least.

As mentioned above, the $\gamma$-ray flux due to the dark matter annihilation 
provide a way to detect the minihalos, such as UCMHs. In Ref.~\cite{Bringmann:2011ut}, 
the authors used the null detection of $\gamma$-ray flux from UCMHs 
to constrain the fraction of UCMHs, and they found the strongest limits 
are $f_{\rm UCMHs} < 10^{-7}$. Besides the contributions on the local 
$\gamma$-ray flux, UCMHs can also provide a extra contribution on the extragalactic 
$\gamma$-ray background. Using the data of extragalactic $\gamma$-ray background 
from the Fermi-LAT, 
the authors of \cite{jcap} found the strongest limits on the 
fraction of UCMHs are $f_{\rm UCMHs} < 10^{-5}$, and these limits are stronger than 
that obtained using the CMB data~\cite{jcap,epjplus}. For the researches on 
UCMHs, one except to find them directly from the present 
observations. In Ref.~\cite{Mirabal:2016huj}, using the third Fermi-LAT sources catalog, 
the authors found that there are nearly 33\% sources (about 34) which remains unassociated. 
The authors of \cite{Schoonenberg:2016aml} found that there are about 
10 dark matter minihalos in the local space. Therefore, as a kind of potential
high energy astrophysical objects, UCMHs would be present within these unassociated 
minihlaos. Another very interesting way of 
finding UCMHs is to investigate the formation of baryonic structures within UCMHs. 
Due to the early formation time of UCMHs, it was pointed by the authors 
of~\cite{ucmhs_1} that the low-mass and low-metallicity 
Pop III stars could be formed in the early 
time within UCMHs and these stars might survive to the present day. Therefore, 
the surveys of low-metallicity stars would provide one important way of 
finding UCMHs indirectly~\cite{Christlieb:2002dz}. On the other hand, 
these Pop III stars can also provide extra photons to ionize the IGM 
and impact the 21cm background signals. It is excepted that 
in the near future the analysis of data 
from e.g. JWST(James Webb Space Telescope)\footnote{http://www.jwst.nasa.gov/}, 
GMT(Giant Magellan Telescope)\footnote{http://www.gmto.org/} 
and TMT(The Thirty-meter Telescope)\footnote{http://www.tmt.org/} can provide a potential possible way of detecting 
UCMHs indirectly.

\section{Acknowledgments}
We thank Dr. B. Yue for very helpful suggestions. 
Yupeng Yang is supported by the National Science
Foundation of China (Grants No.U1404114 No.11505005 and No.11373068).

\newcommand\PR[3]{~Phys. Rept.{\bf ~#1}, #2~(#3)}
\newcommand\NJP[3]{~New J.Phys.{\bf ~#1}, #2~(#3)}
\newcommand\PRD[3]{~Phys. Rev. D{\bf ~#1}, #2~(#3)}
\newcommand\APJ[3]{~Astrophys. J.{\bf ~#1}, #2~ (#3)}
\newcommand\PRL[3]{~Phys. Rev. Lett.{\bf ~#1}, #2~(#3)}
\newcommand\APJS[3]{~Astron. J. Suppl.{\bf ~#1}, #2~(#3)}
\newcommand\EPJP[3]{~Eur. Phy. J. Plus{\bf ~#1}, #2~(#3)}
\newcommand\Nature[3]{~Nature{\bf ~#1}, #2~(#3)}
\newcommand\JCAP[3]{~JCAP{\bf ~#1}, #2~(#3)}
\newcommand\APJL[3]{~Astrophys. J. Lett.{\bf ~#1}, #2~ (#3)}
\newcommand\MNRAS[3]{~MNRAS{\bf ~#1}, #2~(#3)}
\newcommand\PLB[3]{~Phys. Lett. B{\bf ~#1}, #2~(#3)}
\newcommand\AP[3]{~Astropart. Phys.{\bf ~#1}, #2~(#3)}
\newcommand\ARNPS[3]{~Ann. Rev. Nucl. Part. Sci.{\bf ~#1}, #2~(#3)} 
\newcommand\JPCS[3]{~J. Phys.: Conf. Ser.{\bf ~#1}, #2~(#3)}



\begin{thebibliography}{40}
\expandafter\ifx\csname natexlab\endcsname\relax\def\natexlab#1{#1}\fi
\expandafter\ifx\csname bibnamefont\endcsname\relax
  \def\bibnamefont#1{#1}\fi
\expandafter\ifx\csname bibfnamefont\endcsname\relax
  \def\bibfnamefont#1{#1}\fi
\expandafter\ifx\csname citenamefont\endcsname\relax
  \def\citenamefont#1{#1}\fi
\expandafter\ifx\csname url\endcsname\relax
  \def\url#1{\texttt{#1}}\fi
\expandafter\ifx\csname urlprefix\endcsname\relax\def\urlprefix{URL }\fi
\providecommand{\bibinfo}[2]{#2}
\providecommand{\eprint}[2][]{\url{#2}}

\bibitem[{\citenamefont{Furlanetto
  et~al.}(2006{\natexlab{a}})\citenamefont{Furlanetto, Oh, and
  Briggs}}]{21cm_review_1}
\bibinfo{author}{\bibfnamefont{S.}~\bibnamefont{Furlanetto}},
  \bibinfo{author}{\bibfnamefont{S.~P.} \bibnamefont{Oh}}, \bibnamefont{and}
  \bibinfo{author}{\bibfnamefont{F.}~\bibnamefont{Briggs}},
  \bibinfo{journal}{Phys. Rept.} \textbf{\bibinfo{volume}{433}},
  \bibinfo{pages}{181} (\bibinfo{year}{2006}{\natexlab{a}}),
  \eprint{astro-ph/0608032}.

\bibitem[{\citenamefont{Pritchard and
  Loeb}(2012{\natexlab{a}})}]{21cm_review_2}
\bibinfo{author}{\bibfnamefont{J.~R.} \bibnamefont{Pritchard}}
  \bibnamefont{and} \bibinfo{author}{\bibfnamefont{A.}~\bibnamefont{Loeb}},
  \bibinfo{journal}{Reports on Progress in Physics}
  \textbf{\bibinfo{volume}{75}}, \bibinfo{pages}{086901}
  (\bibinfo{year}{2012}{\natexlab{a}}),
  \urlprefix\url{http://stacks.iop.org/0034-4885/75/i=8/a=086901}.

\bibitem[{\citenamefont{Chen and Kamionkowski}(2004)}]{xlchen}
\bibinfo{author}{\bibfnamefont{X.}~\bibnamefont{Chen}} \bibnamefont{and}
  \bibinfo{author}{\bibfnamefont{M.}~\bibnamefont{Kamionkowski}},
  \bibinfo{journal}{Phys. Rev. D} \textbf{\bibinfo{volume}{70}},
  \bibinfo{pages}{043502} (\bibinfo{year}{2004}),
  \urlprefix\url{http://link.aps.org/doi/10.1103/PhysRevD.70.043502}.

\bibitem[{\citenamefont{Zhang et~al.}(2006)\citenamefont{Zhang, Chen, Lei, and
  Si}}]{lezhang}
\bibinfo{author}{\bibfnamefont{L.}~\bibnamefont{Zhang}},
  \bibinfo{author}{\bibfnamefont{X.}~\bibnamefont{Chen}},
  \bibinfo{author}{\bibfnamefont{Y.-A.} \bibnamefont{Lei}}, \bibnamefont{and}
  \bibinfo{author}{\bibfnamefont{Z.-g.} \bibnamefont{Si}},
  \bibinfo{journal}{Phys. Rev. D} \textbf{\bibinfo{volume}{74}},
  \bibinfo{pages}{103519} (\bibinfo{year}{2006}),
  \urlprefix\url{http://link.aps.org/doi/10.1103/PhysRevD.74.103519}.

\bibitem[{\citenamefont{{Vald{\'e}s} et~al.}(2013)\citenamefont{{Vald{\'e}s},
  {Evoli}, {Mesinger}, {Ferrara}, and {Yoshida}}}]{valdes}
\bibinfo{author}{\bibfnamefont{M.}~\bibnamefont{{Vald{\'e}s}}},
  \bibinfo{author}{\bibfnamefont{C.}~\bibnamefont{{Evoli}}},
  \bibinfo{author}{\bibfnamefont{A.}~\bibnamefont{{Mesinger}}},
  \bibinfo{author}{\bibfnamefont{A.}~\bibnamefont{{Ferrara}}},
  \bibnamefont{and}
  \bibinfo{author}{\bibfnamefont{N.}~\bibnamefont{{Yoshida}}},
  \bibinfo{journal}{Monthly Notices of the Royal Astronomical Society}
  \textbf{\bibinfo{volume}{429}}, \bibinfo{pages}{1705} (\bibinfo{year}{2013}),
  \eprint{1209.2120}.

\bibitem[{\citenamefont{Cumberbatch et~al.}(2010)\citenamefont{Cumberbatch,
  Lattanzi, Silk, Lattanzi, and Silk}}]{deltaTb}
\bibinfo{author}{\bibfnamefont{D.~T.} \bibnamefont{Cumberbatch}},
  \bibinfo{author}{\bibfnamefont{M.}~\bibnamefont{Lattanzi}},
  \bibinfo{author}{\bibfnamefont{J.}~\bibnamefont{Silk}},
  \bibinfo{author}{\bibfnamefont{M.}~\bibnamefont{Lattanzi}}, \bibnamefont{and}
  \bibinfo{author}{\bibfnamefont{J.}~\bibnamefont{Silk}},
  \bibinfo{journal}{Phys. Rev.} \textbf{\bibinfo{volume}{D82}},
  \bibinfo{pages}{103508} (\bibinfo{year}{2010}), \eprint{0808.0881}.

\bibitem[{\citenamefont{Natarajan and Schwarz}(2009)}]{Natarajan}
\bibinfo{author}{\bibfnamefont{A.}~\bibnamefont{Natarajan}} \bibnamefont{and}
  \bibinfo{author}{\bibfnamefont{D.~J.} \bibnamefont{Schwarz}},
  \bibinfo{journal}{Phys. Rev. D} \textbf{\bibinfo{volume}{80}},
  \bibinfo{pages}{043529} (\bibinfo{year}{2009}),
  \urlprefix\url{http://link.aps.org/doi/10.1103/PhysRevD.80.043529}.

\bibitem[{\citenamefont{Ricotti and Gould}(2009)}]{ucmhs_1}
\bibinfo{author}{\bibfnamefont{M.}~\bibnamefont{Ricotti}} \bibnamefont{and}
  \bibinfo{author}{\bibfnamefont{A.}~\bibnamefont{Gould}},
  \bibinfo{journal}{Astrophys. J.} \textbf{\bibinfo{volume}{707}},
  \bibinfo{pages}{979} (\bibinfo{year}{2009}), \eprint{0908.0735}.

\bibitem[{\citenamefont{Scott and Sivertsson}(2009)}]{scott_prl}
\bibinfo{author}{\bibfnamefont{P.}~\bibnamefont{Scott}} \bibnamefont{and}
  \bibinfo{author}{\bibfnamefont{S.}~\bibnamefont{Sivertsson}},
  \bibinfo{journal}{Phys. Rev. Lett.} \textbf{\bibinfo{volume}{103}},
  \bibinfo{pages}{211301} (\bibinfo{year}{2009}),
  \urlprefix\url{http://link.aps.org/doi/10.1103/PhysRevLett.103.211301}.

\bibitem[{\citenamefont{Yang et~al.}(2013{\natexlab{a}})\citenamefont{Yang,
  Yang, and Zong}}]{prd_neutrino}
\bibinfo{author}{\bibfnamefont{Y.}~\bibnamefont{Yang}},
  \bibinfo{author}{\bibfnamefont{G.}~\bibnamefont{Yang}}, \bibnamefont{and}
  \bibinfo{author}{\bibfnamefont{H.}~\bibnamefont{Zong}},
  \bibinfo{journal}{Phys. Rev. D} \textbf{\bibinfo{volume}{87}},
  \bibinfo{pages}{103525} (\bibinfo{year}{2013}{\natexlab{a}}),
  \urlprefix\url{http://link.aps.org/doi/10.1103/PhysRevD.87.103525}.

\bibitem[{\citenamefont{Yang et~al.}(2011{\natexlab{a}})\citenamefont{Yang,
  Feng, Huang, Chen, Lu, and Zong}}]{jcap}
\bibinfo{author}{\bibfnamefont{Y.}~\bibnamefont{Yang}},
  \bibinfo{author}{\bibfnamefont{L.}~\bibnamefont{Feng}},
  \bibinfo{author}{\bibfnamefont{X.}~\bibnamefont{Huang}},
  \bibinfo{author}{\bibfnamefont{X.}~\bibnamefont{Chen}},
  \bibinfo{author}{\bibfnamefont{T.}~\bibnamefont{Lu}}, \bibnamefont{and}
  \bibinfo{author}{\bibfnamefont{H.}~\bibnamefont{Zong}},
  \bibinfo{journal}{Journal of Cosmology and Astroparticle Physics}
  \textbf{\bibinfo{volume}{2011}}, \bibinfo{pages}{020}
  (\bibinfo{year}{2011}{\natexlab{a}}),
  \urlprefix\url{http://stacks.iop.org/1475-7516/2011/i=12/a=020}.

\bibitem[{\citenamefont{Yang et~al.}(2013{\natexlab{b}})\citenamefont{Yang,
  Yang, and Zong}}]{epl}
\bibinfo{author}{\bibfnamefont{Y.-P.} \bibnamefont{Yang}},
  \bibinfo{author}{\bibfnamefont{G.-L.} \bibnamefont{Yang}}, \bibnamefont{and}
  \bibinfo{author}{\bibfnamefont{H.-S.} \bibnamefont{Zong}},
  \bibinfo{journal}{EPL (Europhysics Letters)} \textbf{\bibinfo{volume}{101}},
  \bibinfo{pages}{69001} (\bibinfo{year}{2013}{\natexlab{b}}),
  \urlprefix\url{http://stacks.iop.org/0295-5075/101/i=6/a=69001}.

\bibitem[{\citenamefont{Yang}(2014)}]{ijmpa}
\bibinfo{author}{\bibfnamefont{Y.}~\bibnamefont{Yang}},
  \bibinfo{journal}{International Journal of Modern Physics A}
  \textbf{\bibinfo{volume}{29}}, \bibinfo{pages}{1450194}
  (\bibinfo{year}{2014}),
  \eprint{http://www.worldscientific.com/doi/pdf/10.1142/S0217751X14501942},
  \urlprefix\url{http://www.worldscientific.com/doi/abs/10.1142/S0217751X14501942}.

\bibitem[{\citenamefont{Zheng et~al.}(2014)\citenamefont{Zheng, Yang, Li, and
  Zong}}]{raa}
\bibinfo{author}{\bibfnamefont{Y.-L.} \bibnamefont{Zheng}},
  \bibinfo{author}{\bibfnamefont{Y.-P.} \bibnamefont{Yang}},
  \bibinfo{author}{\bibfnamefont{M.-Z.} \bibnamefont{Li}}, \bibnamefont{and}
  \bibinfo{author}{\bibfnamefont{H.-S.} \bibnamefont{Zong}},
  \bibinfo{journal}{Res. Astron. Astrophys.} \textbf{\bibinfo{volume}{14}},
  \bibinfo{pages}{1215} (\bibinfo{year}{2014}), \eprint{1404.0433}.

\bibitem[{\citenamefont{Yang et~al.}(2011{\natexlab{b}})\citenamefont{Yang,
  Huang, Chen, and Zong}}]{prd_1}
\bibinfo{author}{\bibfnamefont{Y.}~\bibnamefont{Yang}},
  \bibinfo{author}{\bibfnamefont{X.}~\bibnamefont{Huang}},
  \bibinfo{author}{\bibfnamefont{X.}~\bibnamefont{Chen}}, \bibnamefont{and}
  \bibinfo{author}{\bibfnamefont{H.}~\bibnamefont{Zong}},
  \bibinfo{journal}{Phys. Rev. D} \textbf{\bibinfo{volume}{84}},
  \bibinfo{pages}{043506} (\bibinfo{year}{2011}{\natexlab{b}}),
  \urlprefix\url{http://link.aps.org/doi/10.1103/PhysRevD.84.043506}.

\bibitem[{\citenamefont{Yang et~al.}(2011{\natexlab{c}})\citenamefont{Yang,
  Chen, Lu, and Zong}}]{epjplus}
\bibinfo{author}{\bibfnamefont{Y.}~\bibnamefont{Yang}},
  \bibinfo{author}{\bibfnamefont{X.}~\bibnamefont{Chen}},
  \bibinfo{author}{\bibfnamefont{T.}~\bibnamefont{Lu}}, \bibnamefont{and}
  \bibinfo{author}{\bibfnamefont{H.}~\bibnamefont{Zong}}, \bibinfo{journal}{The
  European Physical Journal Plus} \textbf{\bibinfo{volume}{126}},
  \bibinfo{eid}{123} (\bibinfo{year}{2011}{\natexlab{c}}),
  \urlprefix\url{http://dx.doi.org/10.1140/epjp/i2011-11123-8}.

\bibitem[{\citenamefont{Zhang}(2011)}]{dongzhang}
\bibinfo{author}{\bibfnamefont{D.}~\bibnamefont{Zhang}},
  \bibinfo{journal}{Monthly Notices of the Royal Astronomical Society}
  \textbf{\bibinfo{volume}{418}}, \bibinfo{pages}{1850} (\bibinfo{year}{2011}),
  \urlprefix\url{http://mnras.oxfordjournals.org/content/418/3/1850.abstract}.

\bibitem[{\citenamefont{Jungman et~al.}(1996)\citenamefont{Jungman,
  Kamionkowski, and Griest}}]{dm_1}
\bibinfo{author}{\bibfnamefont{G.}~\bibnamefont{Jungman}},
  \bibinfo{author}{\bibfnamefont{M.}~\bibnamefont{Kamionkowski}},
  \bibnamefont{and} \bibinfo{author}{\bibfnamefont{K.}~\bibnamefont{Griest}},
  \bibinfo{journal}{Phys. Rept.} \textbf{\bibinfo{volume}{267}},
  \bibinfo{pages}{195} (\bibinfo{year}{1996}), \eprint{hep-ph/9506380}.

\bibitem[{\citenamefont{Bertone et~al.}(2005)\citenamefont{Bertone, Hooper, and
  Silk}}]{dm_2}
\bibinfo{author}{\bibfnamefont{G.}~\bibnamefont{Bertone}},
  \bibinfo{author}{\bibfnamefont{D.}~\bibnamefont{Hooper}}, \bibnamefont{and}
  \bibinfo{author}{\bibfnamefont{J.}~\bibnamefont{Silk}},
  \bibinfo{journal}{Phys. Rept.} \textbf{\bibinfo{volume}{405}},
  \bibinfo{pages}{279} (\bibinfo{year}{2005}), \eprint{hep-ph/0404175}.

\bibitem[{\citenamefont{Hooper and Goodenough}(2011)}]{Hooper}
\bibinfo{author}{\bibfnamefont{D.}~\bibnamefont{Hooper}} \bibnamefont{and}
  \bibinfo{author}{\bibfnamefont{L.}~\bibnamefont{Goodenough}},
  \bibinfo{journal}{Phys. Lett.} \textbf{\bibinfo{volume}{B697}},
  \bibinfo{pages}{412} (\bibinfo{year}{2011}), \eprint{1010.2752}.

\bibitem[{\citenamefont{Bi et~al.}(2013)\citenamefont{Bi, Yin, and Yuan}}]{xjb}
\bibinfo{author}{\bibfnamefont{X.-J.} \bibnamefont{Bi}},
  \bibinfo{author}{\bibfnamefont{P.-F.} \bibnamefont{Yin}}, \bibnamefont{and}
  \bibinfo{author}{\bibfnamefont{Q.}~\bibnamefont{Yuan}},
  \bibinfo{journal}{Front. Phys. China} \textbf{\bibinfo{volume}{8}},
  \bibinfo{pages}{794} (\bibinfo{year}{2013}), \eprint{1409.4590}.

\bibitem[{\citenamefont{Ackermann et~al.}(2015)\citenamefont{Ackermann, Albert,
  Anderson, Atwood, Baldini, Barbiellini, Bastieri, Bechtol, Bellazzini,
  Bissaldi et~al.}}]{PhysRevLett.115.231301}
\bibinfo{author}{\bibfnamefont{M.}~\bibnamefont{Ackermann}},
  \bibinfo{author}{\bibfnamefont{A.}~\bibnamefont{Albert}},
  \bibinfo{author}{\bibfnamefont{B.}~\bibnamefont{Anderson}},
  \bibinfo{author}{\bibfnamefont{W.~B.} \bibnamefont{Atwood}},
  \bibinfo{author}{\bibfnamefont{L.}~\bibnamefont{Baldini}},
  \bibinfo{author}{\bibfnamefont{G.}~\bibnamefont{Barbiellini}},
  \bibinfo{author}{\bibfnamefont{D.}~\bibnamefont{Bastieri}},
  \bibinfo{author}{\bibfnamefont{K.}~\bibnamefont{Bechtol}},
  \bibinfo{author}{\bibfnamefont{R.}~\bibnamefont{Bellazzini}},
  \bibinfo{author}{\bibfnamefont{E.}~\bibnamefont{Bissaldi}},
  \bibnamefont{et~al.} (\bibinfo{collaboration}{The Fermi-LAT Collaboration}),
  \bibinfo{journal}{Phys. Rev. Lett.} \textbf{\bibinfo{volume}{115}},
  \bibinfo{pages}{231301} (\bibinfo{year}{2015}),
  \urlprefix\url{http://link.aps.org/doi/10.1103/PhysRevLett.115.231301}.

\bibitem[{\citenamefont{Abramowski et~al.}(2011)\citenamefont{Abramowski,
  Acero, Aharonian, Akhperjanian, Anton, Barnacka, Barres~de Almeida,
  Bazer-Bachi, Becherini, Becker et~al.}}]{PhysRevLett.106.161301}
\bibinfo{author}{\bibfnamefont{A.}~\bibnamefont{Abramowski}},
  \bibinfo{author}{\bibfnamefont{F.}~\bibnamefont{Acero}},
  \bibinfo{author}{\bibfnamefont{F.}~\bibnamefont{Aharonian}},
  \bibinfo{author}{\bibfnamefont{A.~G.} \bibnamefont{Akhperjanian}},
  \bibinfo{author}{\bibfnamefont{G.}~\bibnamefont{Anton}},
  \bibinfo{author}{\bibfnamefont{A.}~\bibnamefont{Barnacka}},
  \bibinfo{author}{\bibfnamefont{U.}~\bibnamefont{Barres~de Almeida}},
  \bibinfo{author}{\bibfnamefont{A.~R.} \bibnamefont{Bazer-Bachi}},
  \bibinfo{author}{\bibfnamefont{Y.}~\bibnamefont{Becherini}},
  \bibinfo{author}{\bibfnamefont{J.}~\bibnamefont{Becker}},
  \bibnamefont{et~al.} (\bibinfo{collaboration}{H.E.S.S. Collaboration}),
  \bibinfo{journal}{Phys. Rev. Lett.} \textbf{\bibinfo{volume}{106}},
  \bibinfo{pages}{161301} (\bibinfo{year}{2011}),
  \urlprefix\url{http://link.aps.org/doi/10.1103/PhysRevLett.106.161301}.

\bibitem[{\citenamefont{Ade et~al.}(2014)}]{planck}
\bibinfo{author}{\bibfnamefont{P.~A.~R.} \bibnamefont{Ade}}
  \bibnamefont{et~al.} (\bibinfo{collaboration}{Planck}),
  \bibinfo{journal}{Astron. Astrophys.} \textbf{\bibinfo{volume}{571}},
  \bibinfo{pages}{A16} (\bibinfo{year}{2014}), \eprint{1303.5076}.

\bibitem[{\citenamefont{Yuan et~al.}(2010)\citenamefont{Yuan, Yue, Bi, Chen,
  and Zhang}}]{binyue}
\bibinfo{author}{\bibfnamefont{Q.}~\bibnamefont{Yuan}},
  \bibinfo{author}{\bibfnamefont{B.}~\bibnamefont{Yue}},
  \bibinfo{author}{\bibfnamefont{X.-J.} \bibnamefont{Bi}},
  \bibinfo{author}{\bibfnamefont{X.}~\bibnamefont{Chen}}, \bibnamefont{and}
  \bibinfo{author}{\bibfnamefont{X.}~\bibnamefont{Zhang}},
  \bibinfo{journal}{JCAP} \textbf{\bibinfo{volume}{1010}}, \bibinfo{pages}{023}
  (\bibinfo{year}{2010}), \eprint{0912.2504}.

\bibitem[{\citenamefont{Valdes et~al.}(2007)\citenamefont{Valdes, Ferrara,
  Mapelli, and Ripamonti}}]{j_alpha}
\bibinfo{author}{\bibfnamefont{M.}~\bibnamefont{Valdes}},
  \bibinfo{author}{\bibfnamefont{A.}~\bibnamefont{Ferrara}},
  \bibinfo{author}{\bibfnamefont{M.}~\bibnamefont{Mapelli}}, \bibnamefont{and}
  \bibinfo{author}{\bibfnamefont{E.}~\bibnamefont{Ripamonti}},
  \bibinfo{journal}{Mon. Not. Roy. Astron. Soc.}
  \textbf{\bibinfo{volume}{377}}, \bibinfo{pages}{245} (\bibinfo{year}{2007}),
  \eprint{astro-ph/0701301}.

\bibitem[{\citenamefont{Ripamonti et~al.}(2008)\citenamefont{Ripamonti,
  Mapelli, and Zaroubi}}]{Ripamonti}
\bibinfo{author}{\bibfnamefont{E.}~\bibnamefont{Ripamonti}},
  \bibinfo{author}{\bibfnamefont{M.}~\bibnamefont{Mapelli}}, \bibnamefont{and}
  \bibinfo{author}{\bibfnamefont{S.}~\bibnamefont{Zaroubi}},
  \bibinfo{journal}{Mon. Not. Roy. Astron. Soc.}
  \textbf{\bibinfo{volume}{387}}, \bibinfo{pages}{158} (\bibinfo{year}{2008}),
  \eprint{0802.1857}.

\bibitem[{\citenamefont{Kuhlen et~al.}(2006)\citenamefont{Kuhlen, Madau, and
  Montgomery}}]{Kuhlen}
\bibinfo{author}{\bibfnamefont{M.}~\bibnamefont{Kuhlen}},
  \bibinfo{author}{\bibfnamefont{P.}~\bibnamefont{Madau}}, \bibnamefont{and}
  \bibinfo{author}{\bibfnamefont{R.}~\bibnamefont{Montgomery}},
  \bibinfo{journal}{Astrophys. J.} \textbf{\bibinfo{volume}{637}},
  \bibinfo{pages}{L1} (\bibinfo{year}{2006}), \eprint{astro-ph/0510814}.

\bibitem[{\citenamefont{Liszt}(2001)}]{Liszt}
\bibinfo{author}{\bibfnamefont{H.}~\bibnamefont{Liszt}},
  \bibinfo{journal}{Astron. Astrophys.} \textbf{\bibinfo{volume}{371}},
  \bibinfo{pages}{698} (\bibinfo{year}{2001}), \eprint{astro-ph/0103246}.

\bibitem[{\citenamefont{Ciardi and Madau}(2003)}]{Ciardi}
\bibinfo{author}{\bibfnamefont{B.}~\bibnamefont{Ciardi}} \bibnamefont{and}
  \bibinfo{author}{\bibfnamefont{P.}~\bibnamefont{Madau}},
  \bibinfo{journal}{Astrophys. J.} \textbf{\bibinfo{volume}{596}},
  \bibinfo{pages}{1} (\bibinfo{year}{2003}), \eprint{astro-ph/0303249}.

\bibitem[{\citenamefont{Pritchard and Loeb}(2012{\natexlab{b}})}]{review_1}
\bibinfo{author}{\bibfnamefont{J.~R.} \bibnamefont{Pritchard}}
  \bibnamefont{and} \bibinfo{author}{\bibfnamefont{A.}~\bibnamefont{Loeb}},
  \bibinfo{journal}{Reports on Progress in Physics}
  \textbf{\bibinfo{volume}{75}}, \bibinfo{pages}{086901}
  (\bibinfo{year}{2012}{\natexlab{b}}),
  \urlprefix\url{http://stacks.iop.org/0034-4885/75/i=8/a=086901}.

\bibitem[{\citenamefont{Furlanetto
  et~al.}(2006{\natexlab{b}})\citenamefont{Furlanetto, Oh, and
  Briggs}}]{review_2}
\bibinfo{author}{\bibfnamefont{S.}~\bibnamefont{Furlanetto}},
  \bibinfo{author}{\bibfnamefont{S.~P.} \bibnamefont{Oh}}, \bibnamefont{and}
  \bibinfo{author}{\bibfnamefont{F.}~\bibnamefont{Briggs}},
  \bibinfo{journal}{Phys. Rept.} \textbf{\bibinfo{volume}{433}},
  \bibinfo{pages}{181} (\bibinfo{year}{2006}{\natexlab{b}}),
  \eprint{astro-ph/0608032}.

\bibitem[{\citenamefont{Ullio et~al.}(2002)\citenamefont{Ullio, Bergstr\"om,
  Edsj\"o, and Lacey}}]{rcut}
\bibinfo{author}{\bibfnamefont{P.}~\bibnamefont{Ullio}},
  \bibinfo{author}{\bibfnamefont{L.}~\bibnamefont{Bergstr\"om}},
  \bibinfo{author}{\bibfnamefont{J.}~\bibnamefont{Edsj\"o}}, \bibnamefont{and}
  \bibinfo{author}{\bibfnamefont{C.}~\bibnamefont{Lacey}},
  \bibinfo{journal}{Phys. Rev. D} \textbf{\bibinfo{volume}{66}},
  \bibinfo{pages}{123502} (\bibinfo{year}{2002}),
  \urlprefix\url{http://link.aps.org/doi/10.1103/PhysRevD.66.123502}.

\bibitem[{\citenamefont{Green et~al.}(2005)\citenamefont{Green, Hofmann, and
  Schwarz}}]{dmhalos}
\bibinfo{author}{\bibfnamefont{A.~M.} \bibnamefont{Green}},
  \bibinfo{author}{\bibfnamefont{S.}~\bibnamefont{Hofmann}}, \bibnamefont{and}
  \bibinfo{author}{\bibfnamefont{D.~J.} \bibnamefont{Schwarz}},
  \bibinfo{journal}{JCAP} \textbf{\bibinfo{volume}{0508}}, \bibinfo{pages}{003}
  (\bibinfo{year}{2005}), \eprint{astro-ph/0503387}.

\bibitem[{\citenamefont{Diemand et~al.}(2005)\citenamefont{Diemand, Moore, and
  Stadel}}]{subhalos}
\bibinfo{author}{\bibfnamefont{J.}~\bibnamefont{Diemand}},
  \bibinfo{author}{\bibfnamefont{B.}~\bibnamefont{Moore}}, \bibnamefont{and}
  \bibinfo{author}{\bibfnamefont{J.}~\bibnamefont{Stadel}},
  \bibinfo{journal}{Nature} \textbf{\bibinfo{volume}{433}},
  \bibinfo{pages}{389} (\bibinfo{year}{2005}), \eprint{astro-ph/0501589}.

\bibitem[{\citenamefont{Bowman et~al.}(2008)\citenamefont{Bowman, Rogers, and
  Hewitt}}]{0004-637X-676-1-1}
\bibinfo{author}{\bibfnamefont{J.~D.} \bibnamefont{Bowman}},
  \bibinfo{author}{\bibfnamefont{A.~E.~E.} \bibnamefont{Rogers}},
  \bibnamefont{and} \bibinfo{author}{\bibfnamefont{J.~N.}
  \bibnamefont{Hewitt}}, \bibinfo{journal}{The Astrophysical Journal}
  \textbf{\bibinfo{volume}{676}}, \bibinfo{pages}{1} (\bibinfo{year}{2008}),
  \urlprefix\url{http://stacks.iop.org/0004-637X/676/i=1/a=1}.

\bibitem[{\citenamefont{Bringmann et~al.}(2012)\citenamefont{Bringmann, Scott,
  and Akrami}}]{Bringmann:2011ut}
\bibinfo{author}{\bibfnamefont{T.}~\bibnamefont{Bringmann}},
  \bibinfo{author}{\bibfnamefont{P.}~\bibnamefont{Scott}}, \bibnamefont{and}
  \bibinfo{author}{\bibfnamefont{Y.}~\bibnamefont{Akrami}},
  \bibinfo{journal}{Phys. Rev.} \textbf{\bibinfo{volume}{D85}},
  \bibinfo{pages}{125027} (\bibinfo{year}{2012}), \eprint{1110.2484}.

\bibitem[{\citenamefont{Mirabal et~al.}(2016)\citenamefont{Mirabal, Charles,
  Ferrara, Gonthier, Harding, Sánchez-Conde, and Thompson}}]{Mirabal:2016huj}
\bibinfo{author}{\bibfnamefont{N.}~\bibnamefont{Mirabal}},
  \bibinfo{author}{\bibfnamefont{E.}~\bibnamefont{Charles}},
  \bibinfo{author}{\bibfnamefont{E.~C.} \bibnamefont{Ferrara}},
  \bibinfo{author}{\bibfnamefont{P.~L.} \bibnamefont{Gonthier}},
  \bibinfo{author}{\bibfnamefont{A.~K.} \bibnamefont{Harding}},
  \bibinfo{author}{\bibfnamefont{M.~A.} \bibnamefont{Sánchez-Conde}},
  \bibnamefont{and} \bibinfo{author}{\bibfnamefont{D.~J.}
  \bibnamefont{Thompson}}, \bibinfo{journal}{Astrophys. J.}
  \textbf{\bibinfo{volume}{825}}, \bibinfo{pages}{69} (\bibinfo{year}{2016}),
  \eprint{1605.00711}.

\bibitem[{\citenamefont{Schoonenberg et~al.}(2016)\citenamefont{Schoonenberg,
  Gaskins, Bertone, and Diemand}}]{Schoonenberg:2016aml}
\bibinfo{author}{\bibfnamefont{D.}~\bibnamefont{Schoonenberg}},
  \bibinfo{author}{\bibfnamefont{J.}~\bibnamefont{Gaskins}},
  \bibinfo{author}{\bibfnamefont{G.}~\bibnamefont{Bertone}}, \bibnamefont{and}
  \bibinfo{author}{\bibfnamefont{J.}~\bibnamefont{Diemand}},
  \bibinfo{journal}{JCAP} \textbf{\bibinfo{volume}{1605}}, \bibinfo{pages}{028}
  (\bibinfo{year}{2016}), \eprint{1601.06781}.

\bibitem[{\citenamefont{Christlieb et~al.}(2002)\citenamefont{Christlieb,
  Bessell, Beers, Gustafsson, Korn, Barklem, Karlsson, Mizuno-Wiedner, and
  Rossi}}]{Christlieb:2002dz}
\bibinfo{author}{\bibfnamefont{N.}~\bibnamefont{Christlieb}},
  \bibinfo{author}{\bibfnamefont{M.~S.} \bibnamefont{Bessell}},
  \bibinfo{author}{\bibfnamefont{T.~C.} \bibnamefont{Beers}},
  \bibinfo{author}{\bibfnamefont{B.}~\bibnamefont{Gustafsson}},
  \bibinfo{author}{\bibfnamefont{A.~J.} \bibnamefont{Korn}},
  \bibinfo{author}{\bibfnamefont{P.~S.} \bibnamefont{Barklem}},
  \bibinfo{author}{\bibfnamefont{T.}~\bibnamefont{Karlsson}},
  \bibinfo{author}{\bibfnamefont{M.}~\bibnamefont{Mizuno-Wiedner}},
  \bibnamefont{and} \bibinfo{author}{\bibfnamefont{S.}~\bibnamefont{Rossi}},
  \bibinfo{journal}{Nature} \textbf{\bibinfo{volume}{419}},
  \bibinfo{pages}{904} (\bibinfo{year}{2002}), \eprint{astro-ph/0211274}.

\end{thebibliography}

\end{document}